\documentclass[10pt,journal,compsoc]{IEEEtran}
\usepackage{multirow}
\usepackage{xcolor}
\usepackage[normalem]{ulem}
\usepackage{graphicx}
\usepackage{array}
\usepackage{amsmath}
\usepackage{lineno,hyperref}
\usepackage{soul,xcolor}
\usepackage{rotating}
\setstcolor{red}
\DeclareMathAlphabet{\mm}{OT1}{pzc}{m}{it}

\ifCLASSINFOpdf
\else
\fi
\ifCLASSOPTIONcompsoc
  \usepackage[nocompress]{cite}
\else
  \usepackage{cite}
\fi
\newcommand{\note}[1]{{\color{black}  #1}}

\begin{document}
\bstctlcite{IEEEexample:BSTcontrol}
\title{PaRTAA: A Real-time Multiprocessor for Mixed-Criticality Airborne Systems}
%
%
%

\author{Shibarchi~Majumder~\IEEEmembership{Member,~IEEE,}~Jens~F~D~Nielsen and~Thomas~Bak~\IEEEmembership{Senior Member,~IEEE}
         

\thanks{The authors are with the Department of 
Electronic Systems, 
Aalborg University, Aalborg 9220, Denmark. email: 
sm, jdn, tba (@es.aau.dk).}
\thanks{This research is funded by the Danish Independent Research Foundation under grant number 
6111-00363B.}
}

\markboth{Published: IEEE TRANSACTIONS ON COMPUTERS,~Vol.~69, Issue.~8, 01~August~2020 DOI: 10.1109/TC.2020.3002697}%
{Shibarchi \MakeLowercase{\textit{et al.}}: }

\IEEEtitleabstractindextext{%
\begin{abstract}
Mixed-criticality systems, where multiple systems with varying criticality-levels share a single hardware platform, require isolation between tasks with different criticality-levels. Isolation can be achieved with software-based solutions or can be enforced by a hardware level partitioning. An asymmetric multiprocessor architecture offers hardware-based isolation at the cost of underutilized hardware resources, and the inter-core communication mechanism is often a single point of failure in such architectures. In contrast, a partitioned uniprocessor offers efficient resource utilization at the cost of limited scalability.

We propose a partitioned real-time asymmetric architecture (PaRTAA) specifically designed for mixed-criticality airborne systems, featuring robust partitioning within processing elements for establishing isolation between tasks with varying criticality. The granularity in the processing element offers efficient resource utilization where inter-dependent tasks share the same processing element for sequential execution while preserving isolation, and independent tasks simultaneously execute on different processing elements as per system requirements. 

\end{abstract}

\begin{IEEEkeywords}
Avionics on Multi-core, Mixed-criticality Systems, Integrated Modular Avionics, Robust Resource Partitioning, Single Core Equivalence, Processor Architecture.
\end{IEEEkeywords}}

\maketitle

\IEEEdisplaynontitleabstractindextext

%
\IEEEpeerreviewmaketitle

\IEEEraisesectionheading{\section{Introduction}\label{sec:introduction}}
%
%
%
%

\IEEEPARstart{M}{ixed-criticality} systems, where multiple subsystems of different criticality-levels share the same platform, require isolation between subsystems with different criticality-levels to prevent interference that can be addressed by a software-based mechanism such as an operating system or a hypervisor, or can be established at the hardware level. 
Driven by the certification requirements in \textit{safety-critical systems}, e.g., \textit{RTCA DO-297} \cite{DO297} for airborne platforms, the isolation mechanism shall meet the certification requirements of the subsystem with highest criticality-level to be implemented on the platform. Such requirements result in expensive software-based isolation mechanisms and thus enhance the scope of hardware-based isolation. 

\textit{Single-core-equivalent} (SCE) multicore \cite{LuiSha}, an asymmetric multiprocessor (AMP) architecture with multi or many isolated processors with dedicated resources, interconnected over a \textit{network-on-chip} (NoC), offers isolation between tasks running on separate processors. Such an architecture has the potential to meet mixed-criticality requirements for safety-critical airborne platforms where each subsystem can execute on a dedicated processing element without any interference from other subsystems. 
Moreover, an SCE architecture is beneficial over a symmetric multiprocessing architecture in terms of re-usability of existing source-code that avails a more practical and economic transition to multicore for the aerospace industry.
However, in SCE architecture, the NoC or inter-core communication system is a bottleneck and often a single point of failure that requires redundancy or additional safety-net mechanism. Additionally, an one-to-one mapping of individual \textit{integrated modular avionics} (IMA) \cite{DO297} application to AMP processors increases the bandwidth requirement resulting in resource-extensive NoC architecture and adds to the complexity of system scheduling due to inter-core communication delays over the NoC. Furthermore, dedicating a core to a subsystem is not efficient in terms of resource utilization; large airborne platforms may afford many-core platforms with hundreds of cores, but that may not be feasible for resource-limited platforms like \textit{unmanned aerial systems} (UAS).  

Alternatively, a single-core architecture with a hardware-level isolation mechanism can offer better computational resource utilization with a small hardware footprint. However, a partitioned single-core architecture may not be sufficient to address the growing demand for computational resources with an increasing degree of autonomous flight capabilities. 

The SCE architecture offers an excellent solution for parallel execution at the cost of under-utilized resources where a partitioned architecture provides sequential access to a single processing element resulting in efficient resource utilization. 
Aerospace systems have limited scope of parallel execution as some systems may have inter-system dependencies and must be executed sequentially, and mutually independent systems can be executed simultaneously.

\begin{figure}[h!]
\begin{center}
\includegraphics[scale= 0.58]{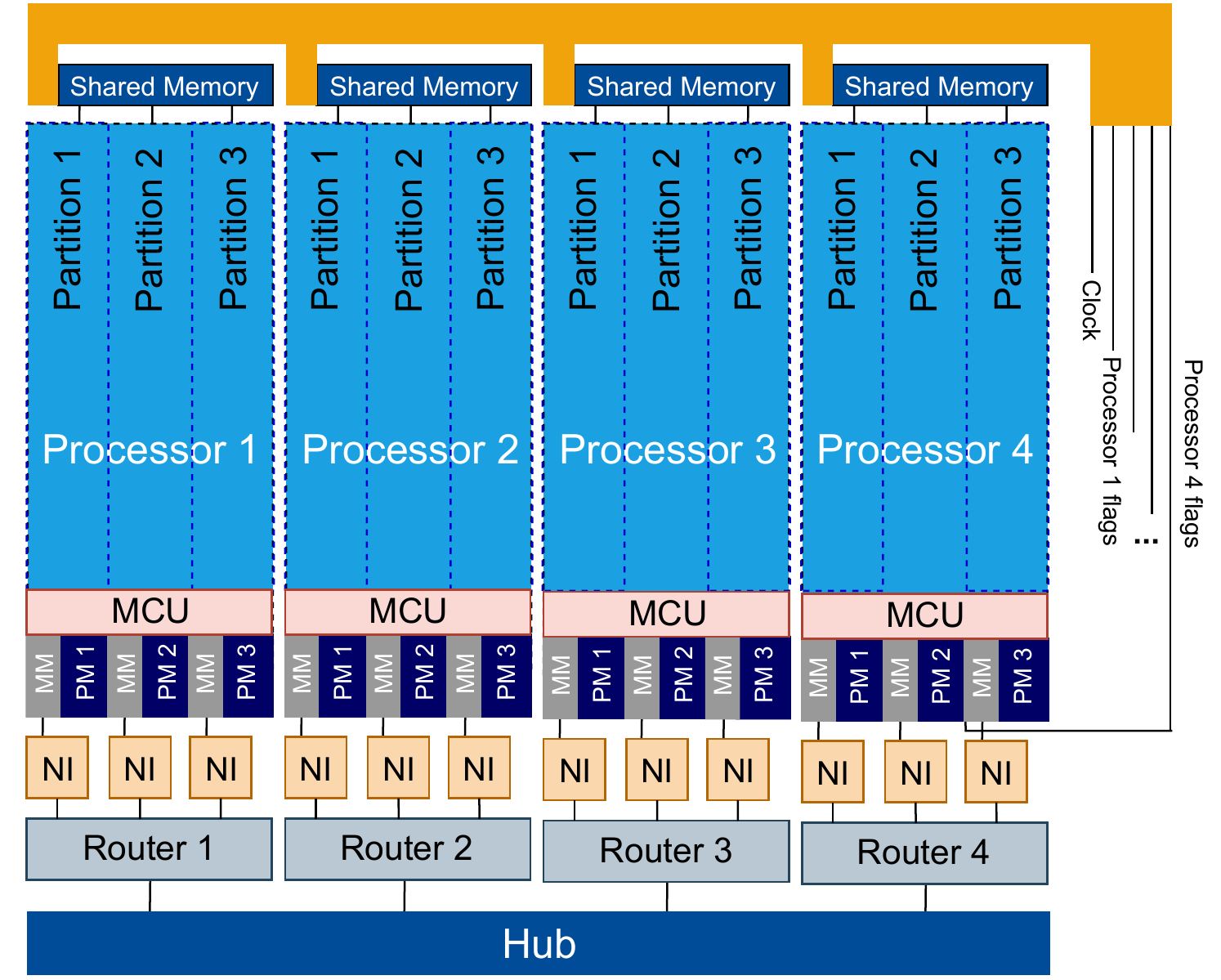}
\caption{A block diagram of the proposed multiprocessor architecture PaRTAA with four processors. Each processor has three underlying partitions with dedicated resources and protected memories (PM). All the partitions are interconnected over a NoC via network interfaces(NI), memory-mapped (MM) to the protected memory region.}
\label{fig:mpu}
\end{center}
\end{figure}

In this work, we propose a partitioned real-time asymmetric architecture \textit{PaRTAA}, where individual processing elements featuring robust resource partitioning are interconnected with a time predictable NoC to provide a time analyzable resource-partitioned distributed architecture, as shown in Figure \ref{fig:mpu}. The partitions within processing elements satisfy isolation requirements for mixed-criticality systems; sequential tasks can execute on the same or different partitions depending on its criticality-level on the same processor, and independent tasks/ systems can execute simultaneously on separate processors, resulting in efficient resource utilization. 

In our previous work, we proposed a coarse-grained uniprocessor, featuring hardware-level partitioning for isolation \cite{aero}. In another work, we introduced a network-on-chip that offers time analyzability and end-to-end isolation between data packets \cite{majumder_nielsen} for mixed-criticality systems. In this work, we have extended our previous works to propose a multiprocessor architecture featuring hardware-level robust resource partitioning.
The specific contribution of this paper includes - 
\begin{itemize}
    \item A time-predictable multiprocessor architecture featuring robust resource partitioning to accommodate systems with mixed-criticality while preventing interference.
    \item A multiprocessor architecture with coarse-grained processing elements for efficient resource utilization by combining sequential and parallel execution depending upon system requirements.
    \item The architecture preserves the reusability of software solutions in airborne systems developed for single-core platforms.
    \item A globally visible processor flag mechanism for task synchronization under different partitions and processors.
    \item A demonstration of the proposed architecture for an avionics use case.
    
\end{itemize}

\section{Background and Related Work}
The scope of multicore architecture in airborne systems is still evolving. However, several researchers have addressed multicore architectures for real-time and safety-critical applications. This section discusses related architectures in the context of mixed-criticality airborne systems. 

\subsection{Mixed-Criticality and IMA}
Driven by the constraints of space, weight and power, mixed-criticality systems allow systems with different criticality-levels to share the same host hardware when inter-system interference is mitigated. In airborne systems, 
to replace federated avionics architecture, a \textit{line replaceable unit} (LRU) based framework where a single LRU is certified to a single DO-178B \cite{DO178B} and/ or DO-254 \cite{DO254} level, the concept of mixed-criticality system is adopted as IMA \cite{DO297}, where systems with different level of criticality defined as \textit{design assurance level} (DAL) A, B, C, D and E \cite{DO178B, DO178C} can share the same host platform. 
To establish isolation between systems with different criticality-levels, the guidelines for \textit{robust resource partitioning} (RRP)\cite{CAST32A} is adopted from DO-248C \cite{DO248C} and DO-297\cite{DO297}. 

RRP is achieved when: 1) partitions cannot contaminate instructions, I/O, and data of other partitions, 2) partitions cannot use resource outside its allocated resources and 3) dysfunction/ failure in one partition cannot cause adverse effect on other partitions \cite{CAST32A}. 

In practice, IMA system software is separated from the host platform by an \textit{application executive} (APEX), an application programming interface (API) defined in ARINC 653 \cite{ARINC653} standard for IMA partitioning. The ARINC 653 API is independent from the base hardware platform and a given software host environment or operating system implementation. The hardware-software isolation allows incremental certification, thus, modification in systems can be certified by certifying only the changes.  

\subsection{Related Processor Architecture}
Currently, Federation Aviation Administration (FAA) guidelines do not include the use of commercial off the shelf (COTS) microprocessors and system-on-chips (SoCs) in airborne system, however, the scope is under evaluation \cite{Mahapatra2011MicroprocessorEF}\cite{Mahapatra2009MicroprocessorEF}. Furthermore, FAA restricts the use to only one active core in a multicore platform due to potential inter-core interference \cite{CAST32A}.

According to a recent position paper on multicore processors (MCP) for airborne systems \cite{CAST32A}, the applicant should identify all potential interference paths on an MCP for DAL level A and level B applications. 
When implemented on an MCP platform without a robust partitioning, software components or sets of requirements for which interference paths are not avoided or mitigated shall be verified with all other software components executing all together \cite{CAST32A}. However, for MCP platforms with established robust partitioning, software components can be verified separately, including determination of \textit{worst case execution time} (WCET) \cite{CAST32A}. 

The concept of interconnecting multiple partitioned processors to form a distributed partitioned system was proposed in a paper by NASA in 2000 \cite{rushby2000partitioning}. The architecture was found to be challenging due to requirements of point-to-point communication mechanism for inter-partition communication that cannot be achieved with shared bus. However, the feasibility of such an architecture with a data-concentrator unit to hold partition specific data was realized. In our approach, the proposed NoC architecture features similar dedicated sampling buffer to hold partition specific data without interrupting executing partitions.

The idea of mapping RTOS functionalities to hardware is not new. In FASTCHART \cite{144077}, the authors have achieved time deterministic execution with 64 different tasks with 8 different priorities with a RISC architecture by removing cache and pipeline. Adomat et al., in \cite{557849} demonstrated a real-time hardware kernel that can support a similar number of tasks and priorities like FASTCHART, but with additional hardware features.
Ungerer et al. \cite{5567091}, introduced a worst-case time-analyzable multicore architecture for mixed-criticality system with isolation between tasks, although, the architecture is limited to a single hard real-time task per core.

In recent years, Zimmer et al. \cite{6925994} has introduced FlexPRET, a fine-grained multi-threaded processor architecture for mixed-criticality systems. The work, demonstrates a WCET analyzable framework that features isolation in temporal and spatial domain without wasting computational cycles. Tasks are segregated in threads and each thread is given access to the computational resources for a single clock cycle in a fixed or active round-robin arbitration routine. 
A very similar architecture \cite{6378622} by Liu et al., PTARM, is a fine-grained multi-threaded architecture with fixed-round-robin arbitration among four threads. The architecture avails hard real-time execution time at the cost of wasted cycles when all the threads are not active. 
Another similar architecture XMOS \cite{6341002}, poses better utilization of resources by excluding inactive tasks from the scheduling. WCET is analyzed by considering the maximum number of active tasks at any given time. 
Delvai et al. \cite{1212740}, introduced a 16-bit, 3 stage processor, SPEAR, with repeatable-time instructions by single path execution flow.

Apart from the partitioned uniprocessor architectures, there are several multiprocessor architectures proposed for generic mixed-criticality systems. 
Salloum et al. in \cite{ELSALLOUM20131020}, has proposed and demonstrated a multiprocessor system on chip (MPSoC) solution that focuses on easing the certification process required in safety-critical domain. MultiPARTES project \cite{TRUJILLO2014921} demonstrates a multi-core heterogeneous architecture and focus on associated tool-chains for easing the certification process for mixed-criticality systems on a multi-core syste.
The researchers in \cite{PEREZ2017145}, describe a system architecture that enables the use of data-centric distribution middleware in partitioned real-time embedded systems based on a hypervisor for multi-core focused on the analysis of the available architectural configurations.
For more related work on mixed-criticality system on multiprocessor, please consider the report by Baruah et al. \cite{baruah2015mixed} that summarizes a broad spectrum of researches related to multi-core and 
many-core architectures in the context of mixed-criticality implementation.
\subsection{Inter-Core Communication}
\note{Several NoC architectures have been demonstrated for safety critical systems. The use of NoC in a real-time system imposes complex constraints in the overall design \cite{Sano2015}. 

In SoCBUS \cite{16}, a \textit{circuit switching} method is applied and a concept called packet connected circuit was introduced, where a data packet propagates through a dynamic minimum route, locking the circuit as it moves through the network. For such switching mechanism, a real-time analysis is possible when the traffic follows a fixed schedule, but not effective where the transmission sequence is not fixed like in avionics systems, where the data sequence depend on the relative state of the applications. 
In Xpipes \cite{15}, the network is tailored to meet the bandwidth requirements of the payload application at its design stage. In practice, such a system could be hard to fabricate as foreseeing the exact communication load is difficult to analyze, and it restricts reusability and the scope of future modification of the system. 
An alternative solution is proposed based on \textit{backtrack probing} to avoid waiting for blocked channels to become available, seeking for alternative non-minimal routes in \cite{18}. However, such a solution does not guarantee time-predictability. 
In \cite{17}, a synchronous circuit switching NoC is presented. A concept of spatial division multiplexing is introduced, where the physical channel is segmented to provide physical separation between data streams at the cost of elevated resources.}

A connection-less packet-switching approach is demonstrated in \cite{26}, where the routers work independently and a \textit{wormhole switching} technique is used. The transmissions are prioritized based on some fixed parameters and transmission with the highest priority is given preference. The draw-back of such a design is that packets with low priorities may be dropped or stalled for a long time and has a longer latency resulting in an unachievable time-analyzability. The authors propose a low end-to-end latency with a \textit{guaranteed service} traffic in \cite{27}. In \cite{28, 29}, the researchers have addressed the low priority packet blocking problem in a connection-less NoC by introducing the concept of increasing priority over waiting time. In contrast, this work offers a mixed, \textit{best-effort} and \textit{guaranteed-service} traffic where transmission with highest priority is given preference by allocating more bandwidth, while transmission with lower priority is given the minimum bandwidth allocated by the system designer to maintain worst-case-time analyzable communication for transmissions belonging to all priorities.


\subsection{Multiprocessor in Avionics}
\begin{figure*}[h!]
\begin{center}
\includegraphics[scale= 0.55]{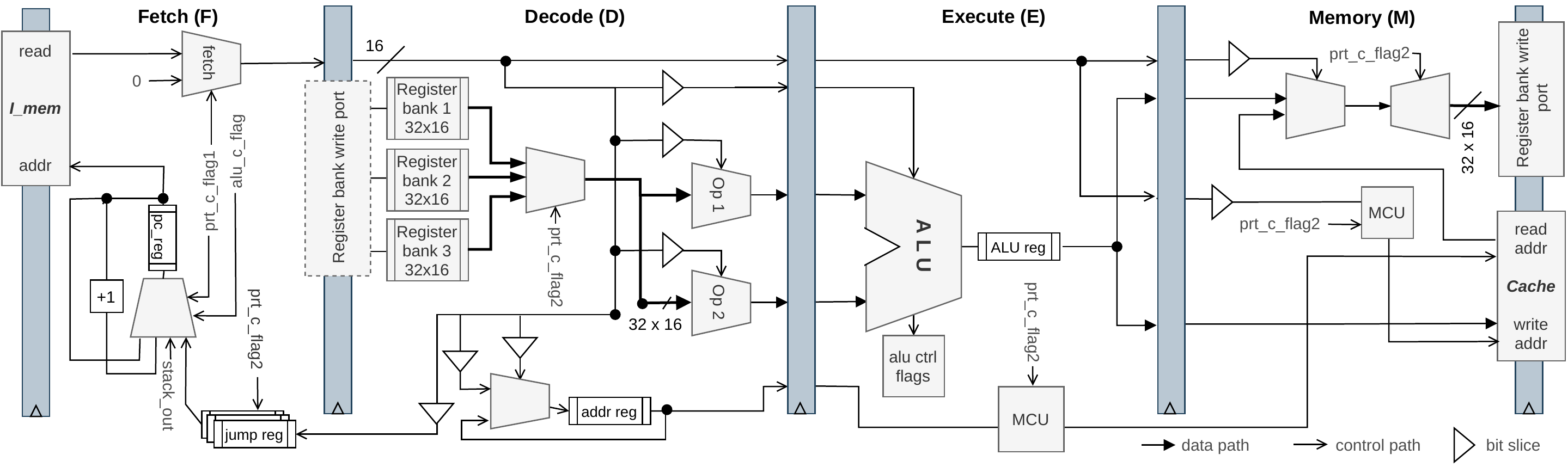}
\caption{A high-level diagram showing the \AE r\o \space processor architecture and pipeline stages with three partitions. Note that the data paths and control paths are specified with different arrows.}
\label{fig:cpu}
\end{center}
\end{figure*} 

System certification with a multicore platform is presently challenging due to lack of guidelines and formal methods issued by certification authorities. The EASA research report \cite{MULCORS} and FAA position paper \cite{CAST32A} are the most recent studies available from a certification authority on the potential scope and regulations on multicore platforms for airborne systems. 
Although, CAST 32A position paper \cite{CAST32A} only includes multicore platforms with only two core and excludes any asymmetric multiprocessors with no shared memory/ cache or coherency module \cite{CAST32A}, SCE architectures are found to be beneficial for mixed-criticality and IMA implementations \cite{LuiSha}. One major benefit of SCE is reusability of source code developed for single core architecture. 

A few AMP and many-core architectures have been demonstrated for IMA implementation. In \cite{Lo}, a helicopter health monitoring system is demonstrated on a many-core processor architecture for exploiting computing capabilities and timing responses. The researchers have used the Kalray MMPA-256 platform, that features dedicated memory banks for processing elements and a reservation mechanism on the NoC.
In \cite{SCHOEBERL2015449}, the authors have proposed a time predictable asymmetric multicore architecture with dual issue real-time processors connected over a time-predictable bi-torus NoC. Finally, a platform with 4 processors has been demonstrated for an avionics application in \cite{7445422}, where each processor hosts a single sub-system of a helicopter health monitoring system.

\section{The Processor}
The proposed multiprocessor architecture is an adoption and extension of our previous work on a uniprocessor, \AE r\o \space \cite{aero}, and we will briefly discuss the essential parts that are required for understanding, and focus on the extensions made in this work. 

\AE r\o \space is a coarse-grained time analyzable processor with a 32-bit RISC style custom ISA and four pipeline stages; instruction fetch (F), decode (D), execute (E), memory-access (M), as shown in Figure \ref{fig:cpu}. \AE r\o \space is statically scheduled, and all delays are analyzable in the instruction set for a single path execution paradigm.

In this work, to fabricate an asymmetric multiprocssor, we have integrated four \AE r\o \space processors, interfaced with a time-predictable NoC, as illustrated in Figure \ref{fig:mpu}. The goal is, a set of tasks of mixed-criticality levels can be scheduled on different partitions under different processors depending on the timing requirements and criticality-level, and the hardware-defined partitioning will feature isolation between tasks with different criticality-levels without any host software support. Please note that the proposed AMP architecture is scalable, and can be fabricated with more processors. That is one of the reasons for considering a NoC instead of an on-chip bus for inter-processor communication.





\subsection{Architecture and Partitioning}
The \AE r\o \space processor features hardware-level partitioning to avail isolation between tasks of different criticality-levels in either temporal or spatial domain. 
Isolation in the spatial domain is achieved with replicated resources, and isolation in the temporal domain is achieved with a partition switching mechanism, Switching-Control-Unit (SwCU). 

The \AE r\o \space offers three partitions to accommodate tasks with three criticality-levels. Each partition has a parallel data and control lines with dedicated and replicated resources connected over a multiplexer. 
Partitioning is a feature built within the processor, and the ISA has no access to it. Furthermore, the instructions have no sense of the partitioning or the partition it is executing within. All the partitions are identical, and any partition can be used to execute instructions of any level of criticality.

As the instruction memory is read-only, a shared memory device can be used to store instructions/ tasks belonging to different criticality-levels. However, once the instruction is fetched, isolation must be maintained (either in the temporal or spatial domain) through the pipeline flow to prevent interference. To feature isolation in data and control flow, parts of the control path and data path contains parallel paths connected over a multiplexer. At any point of operation, only a single path is activated by the SwCU.  

The SwCU is inbuilt hardware that enforces a cycle-accurate time-triggered arbitration mechanism to implement a fixed scheduling configured by the system designer for each partition. The SwCU cannot be reached by the tasks within the processor.  

It is possible to configure different execution period and execution time for each partition in the SwCu, however, to facilitate time predictable execution the following assumptions are made -

\begin{itemize}
    \item \textit{All partitions are periodic.}
    \item \textit{All switching events are time-triggered.}
    \item \textit{All partitions have uniform priority.}
\end{itemize}
{Note that uniform priority suggests that any two partitions shall never compete for execution access and not partitions with same criticality.}

When a partition is active, only the associated dedicated resources (control registers, register bank, and paths) are active, and the instruction flow takes place similar to a non-partitioned processor. When the active partition completes execution access time assigned by the user, the partition is switched, and the next partition (based on partition schedule) is given execution access. All the data and instruction from the inactive partition remain frozen in the respective pipeline/ hardware components (e.g., register banks).

\subsection{Memory Hierarchy}
We have used three types of memories; instruction memory, data memory, and stack memory. 
The instruction memory is read-only memory for the processor, and instructions from different partitions can share the same memory device. We have assumed that all the instructions are present in the instruction memory, and cache misses are out of the scope of this work.

\begin{figure}[h]
\begin{center}
\includegraphics[scale= 0.68]{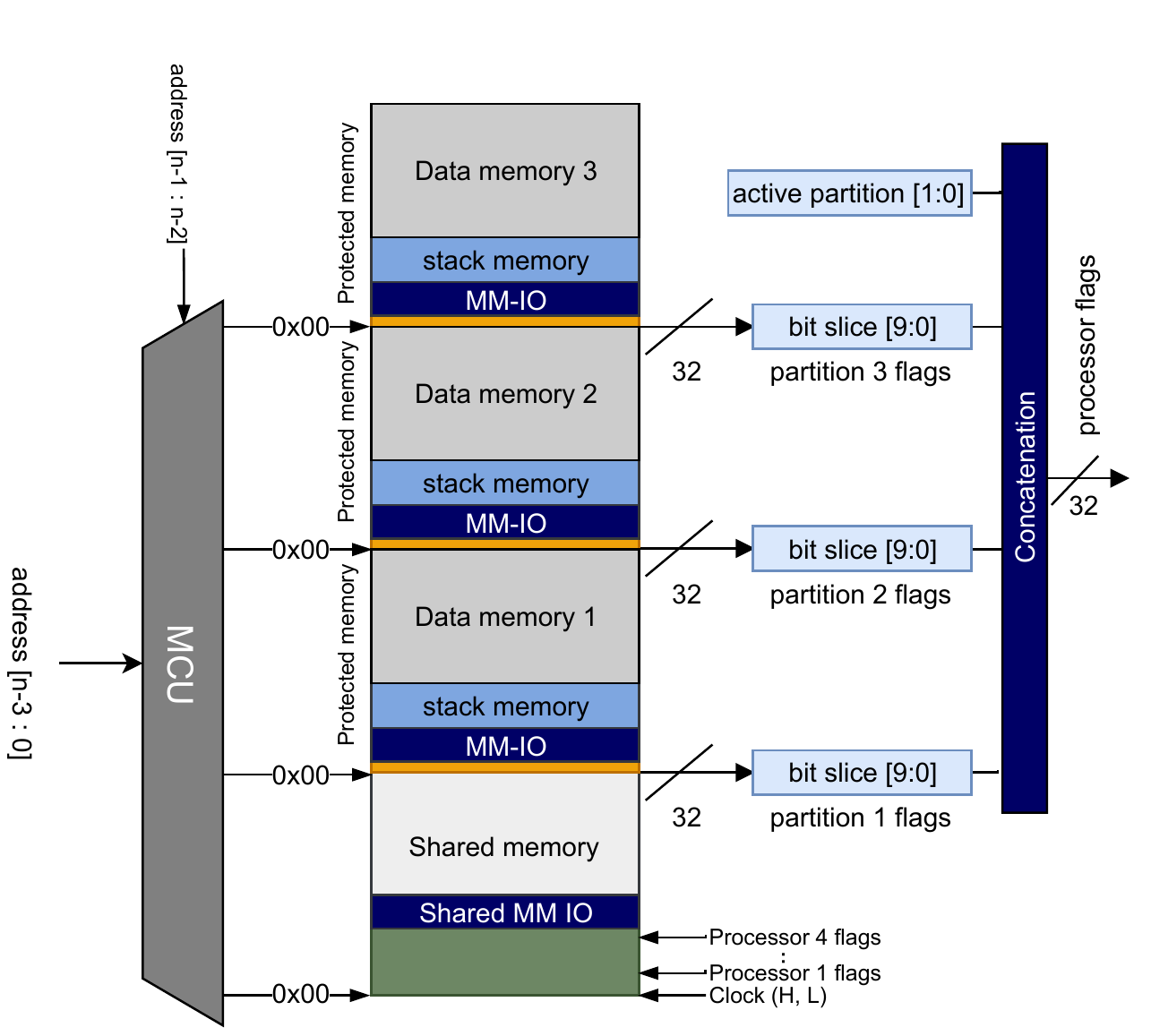}
\caption{Block diagram showing the memory control unit and distribution of data memory of a single processing element.}
\label{fig:mem}
\end{center}
\end{figure} 

The partitions must have read/ write access to the data memory and stack memory for operation, and isolation must be implemented. However, as at any point in time, only a single partition is active, a shared memory device can be used when an isolation mechanism is implemented. A \textit{memory-control-unit} (MCU) establishes such an isolation by partitioning the memory in four segments, as shown in Figure \ref{fig:mem}. Three segments are dedicated to three partitions, and one segment is shared between all three partitions.
The partition specific protected memory segment is used as data memory and stack memory. The shared memory can be used for sharing data between partitions.





The MCU is a hardware-defined mechanism that controls two most-significant-bits (MSB) of the memory address based on the active partition. If the address space of the memory device is $n$-bit, then the MCU controls the $[n-1, n-2]$ -bits of the address space.
The memory address space visible to each partition is $n-1$-bit, i.e., an address space of $[n-2,0]$-bits. However, the $n-2$th bit is used to denote the dedicated or shared memory region.  

The instructions within a partition have no notion of the MCU or the memory segmentation, and the instructions can only select between shared or protected memory region for reading or writing operations.
The MCU implements the following logic in hardware: \def\code#1{\texttt{#1}} \\

\indent \code{\textbf{if} (addr[n-2] == 1):\\
\indent \indent  addr[n-1:n-2] = active-partition-flag;\\
\indent \textbf{elseif} (addr[n-2] == 0):\\
\indent \indent addr[n-1:n-2] = 2b00;}\\

The partition specific dedicated stack is defined on the lower end (except the base address) of the dedicated data memory, and operated by the second access channel of a dual-port memory device. The stack output is set to the most recent \textit{return} address as default. 

At any point of operation, there is only one partition active in a processor; hence, no simultaneous read-write protection is required. However, the shared memory region does not offer any hardware-based protection mechanism, and must have software-based memory protection, e.g., \textit{mutex}.

\subsection{I/O and Peripherals}
The partitioning in the processor does not have any specific requirements for IO interfacing. However, to meet the isolation requirements for mixed-criticality applications, some specific adjustments are necessary. Each partition can be interfaced with external I/O devices with standard \textit{memory-mapping} technique. The I/Os that require isolation is mapped to the protected data memory address space and can only be accessed by the associated partition, and the shared I/Os or peripherals are mapped in the shared address space. 

For broadcasting the execution state within a partition, a custom processor flag mechanism is implemented. The 32-bit processor flag accommodates a 2-bit flag that shows the active partition followed by three 10-bit flags from three partitions. The 10-bit flagging mechanism can be used as 1023 customs user-defined flags.  
The base address of the protected memory region associated with each partition is used to set the flag with a standard memory writing technique, and an underlying hardware mechanism slices the written data and forwards the 10-LSBs, as shown in Figure \ref{fig:mem}.
The processor flags are mapped on the shared data memory space, as presented in Table \ref{Tab:attachments}. The 10-bit partition flag can only be set from the associated partition by accessing the mapped address in the protected data memory region, and all the processor flags can be read by reading the address mapped in the shared address space, as shown in Figure \ref{fig:mem}. The flagging mechanism can be used for execution synchronization, and explained later.


Additionally, the proposed multicore architecture features a global hardware clock accurate to the processor clock cycle. The 64-bit counter is mapped to the base address of the shared memory region. Note that the 64-bit counter has an extremely long cycle time and cannot virtually reset withing runtime. The Table \ref{Tab:attachments} shows the assistive hardware components mapped in the shared memory region. The memory distribution and memory-mapped I/O interfacing are presented in Figure \ref{fig:mem}. 

\renewcommand{\arraystretch}{1.4}
\begin{table}[h]
\caption{Mapping of assistive hardwares in the shared address space.}
\label{Tab:attachments}
\begin{center}
		\begin{tabular}{cc}
			\hline \hline 
			{\textit{Memory-mapped address} }                  & {\textit{attachments}}         \\ \hline 
			\multicolumn{1}{c|}{\code{0x00}} & \code{clock\_L} \textit{(low-bits)}  \\ \hline
			\multicolumn{1}{c|}{\code{0x04}} & \code{clock\_H} \textit{(high-bits)}\\ \hline
			\multicolumn{1}{c|}{\code{0x08}} & \code{Processor\_1\_flags} \textit{(32-bit)} \\ \hline
			\multicolumn{1}{c|}{\code{0x12}} & \code{Processor\_2\_flags} \textit{(32-bit)}\\ \hline
			\multicolumn{1}{c|}{\code{0x16}} & \code{Processor\_3\_flags} \textit{(32-bit)}\\ \hline
			\multicolumn{1}{c|}{\code{0x20}} & \code{Processor\_4\_flags} \textit{(32-bit)}\\ \hline
		\end{tabular}%
\end{center}
\end{table}
\renewcommand{\arraystretch}{1}

\section{Inter-core communication}
Inter-processor connectivity is the basis of asymmetric multiprocessors. For mixed-criticality airborne systems, such an inter-processor communication mechanism must provide time-predictability for real-time applications as well as satisfy isolation between data from tasks with different criticality-levels. To interconnect four processors, we have implemented a time predictable NoC \cite{majumder_nielsen} that offers isolation, either in spatial or in the temporal domain, between data-packets through the entire propagation path as well as it avails data protection from erroneous transmission conducted by potential erroneous systems.

\begin{figure}[h!]
\begin{center}
\includegraphics[scale= 0.6]{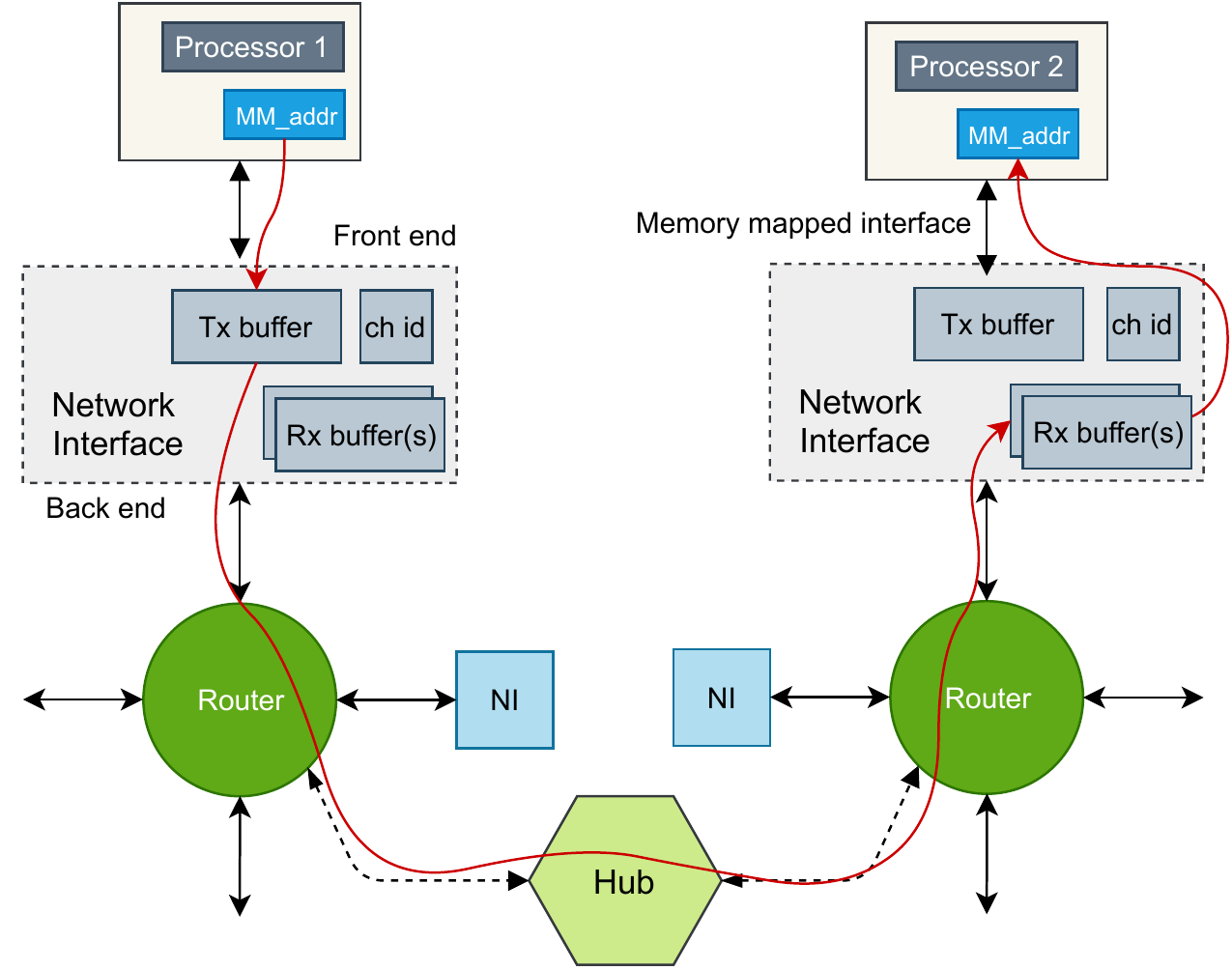}
\caption{Block diagram showing end-to-end data packet transmission.}
\label{fig:flow}
\end{center}
\end{figure}

For easy monitoring and control of network traffic and simpler interfacing, we adopted a hybrid of \textit{star} and \textit{tree} topologies. The network is built around a \textit{hub}, interfaced with multiple \textit{routers} in a star topology and each router is attached to a single or multiple \textit{network-interfaces} (NI) in a reverse tree topology, as shown in Figure \ref{fig:flow}.

Further, to enforce \textit{best-effort} and \textit{guaranteed-service} traffic without any packet loss under normal operation, a mixed arbitration scheme of \textit{time-division multiplexing}(TDM) and \textit{dynamic-priority-token-passing} is proposed.  

The proposed arbitration mechanism accommodates different \textit{latency} and \textit{bandwidth} requirements of different producers/ consumers in the same network while preserving time analyzable end-to-end communication.



\subsection{Interfacing and Propagation}
In PaRTAA, the NoC is configured with four routers, and three NIs connected to each router, as shown in Figure \ref{fig:mpu}. 
The NI has two ends; front-end, to interface with the producer/ consumer and back-end, to interface with the associated router. An NI is independently interfaced with each partition using a standard memory-mapping technique, mapped to the protected memory space of the associated partition. Each NI has a dedicated sampling buffer for each reception channel and each sampling buffer can be separately mapped to the associated partition. However, as a partition can only send a single packet at any point, the NI features a single transmission buffer mapped to the dedicated address space. 

Each partition has a dedicated NI with sampling buffers that protect the received data until consumed by the partition, even if the partition does not have execution access at the time of data reception. A producer can send a data packet any time irrespective of the state of the consumer partition (active or inactive).
The data-packet scheduling considers the number of transmission per unit time, but the exact time of transmission is not fixed, and a producer can push a data packet at any time. A transmission agreement is only violated if the transmission rate exceeds the bandwidth assigned to the channel.

There is no direct memory access (DMA) used in the network, and the producer pushes a data-packet in the network by writing the data-packet into the NI. 
The NI features a transmission sample buffer to temporarily hold the newly written data-packet for one clock cycle. In the subsequent clock cycle, the data packet is forwarded to the router, packed with the destination address in its header. 
As the producer can only write a single data-packet at a time, and the delay between two subsequent writes is more than two clock cycles, temporal isolation between transmission packets is achieved.

At the receiving NI end, a dedicated sampling buffer is featured for each channel to store received data packets. The channel-specific dedicated sampling buffers establish spatial isolation between data-packets from different channels. 

If the production rate is higher than the consumption rate, the old data-packet gets overwritten by a fresh data-packet from the same channel. 
This is not a challenge in the intended application as each sampling buffer holds a complete sample of a specific signal (e.g., turbine temperature, fuel pressure, etc.), and most recent data is desired for computation. 
Similarly, if the transmission rate is lower than the consumption rate, the consumer reads the same data-packet multiple times, and this can be handled in software.

\subsection{Communication Time Analyzability}
In mixed-criticality systems, systems with different levels of criticality can have different timing characteristics. Based on its functionality, a system may have different latency and bandwidth requirements. To accommodate such requirements of different latency and bandwidth in the same network while preserving time-predictability for all communications, a concept of dynamic priority is implemented in the arbitration scheme \cite{majumder_nielsen}.  

Furthermore, communication delay analysis is critical for scheduling of tasks on an asymmetric architecture. The release time of a dependent task (that takes input from other tasks) is influenced by the end-to-end communication delay. Such end-to-end communication delay can be analyzed from the transmission schedule, which is again dependent on task scheduling, creating a deadlock.

In PaRTAA, the propagation time from a producer to the transmission-router via an NI, and reception-router to the destination NI is fixed and equal to 8 clock cycles. The worst-case latency for a router to router propagation via the hub for a single packet transmission can be calculated as $L_{channel} = (\lfloor{S_{total}-1}/{S_{channel}} + 1) \times t_{slot} + 1 $
, where, $L_{channel}$ is the worst-case latency in clock cycles, $S_{total}$ is the total number of slots in the arbitration cycle, and $S_{channel}$ is the number of slots assigned to a channel. 

The worst-case end-to-end communication delay for a specific channel is driven by the slots assigned to the channel and the total number of slots present in the arbitration model, and know before scheduling time.


\section{System Development and Analysis}
The system development approach for an asymmetric multiprocessor is different from the approach taken in more common symmetric multiprocessors. In asymmetric multiprocessing, tasks are strategically segregated among different processors based on respective timing requirements. The robust resource partitioning adds additional complexities as the tasks need to be segregated in different partitions under different processors considering partition switching overheads and inter-task communication delays.
Moreover, apart from WCET, periodicity is crucial for real-time control applications and different systems may require different periodicity that should be accommodated by the partition schedule. 


\subsection{Timing Analysis}
Timing accuracy is critical in airborne systems, and a delayed computation is considered as dysfunction. The timing analysis of a system is more complicated than just analyzing the WCET of the executable binary, and must be analyzed at the system level. The response time of a system depends on the overall timing behavior of individual tasks as well as the characteristics of the integration, including potential delays from inter-tasks dependencies, communication delays, and switching overheads.
\vspace{2mm}

\subsubsection{\textbf{WCET of independent tasks}} Due to the absence of dynamic delays, the processor features a cycle-accurate timing analysis for all tasks. As the processor consumes one instruction in each cycle and each instruction interleaves in the processor pipeline for a fixed number of cycles, the \textit{bounded loop longest path} execution time (in cycles) of a given task $a$, $\tau_a$, is cycle-accurate, repeatable and can be calculated from the executable instructions when executed on a non-partitioned environment (e.g., only one active partition). 

Although a task developed for a single core system does not have the sense of resource partitioning and can directly execute within any partition in the proposed architecture, the partition-scheduling can significantly influence the timing behavior of the underlying tasks.
If a partition $\mm P$, is assigned an execution time $\tau_{p}$ and a periodicity $\lambda_{p}$, and the task $a$ is ported in the partition $\mm P$, then the WCET of task $a$ in partition $\mm P$, $\tau_{p_a}$, can be determined as:
\begin{multline*}
    \tau_{p_a} = \bigg(\bigg\lceil{\frac{\tau_{a}}{\tau_{p}}}\bigg\rceil-1\bigg) \times \lambda_{p} +  \bigg\{ \tau_{a} -  \bigg(\bigg\lceil{\frac{\tau_{a}}{\tau_{p}}}\bigg\rceil - 1\bigg) \times \tau_{p} \bigg\}
\end{multline*}
Note that the partitioning has no effect on the WCET of the task, when $\tau_{p} > \tau_{a}$.
Similarly, if a partition is shared by multiple tasks say $a,b$ and $c$, then the WCET time of individual tasks within the partition $\tau_{p_a}$, $\tau_{p_b}$ and $\tau_{p_c}$ remains unchanged as $\tau_a$, $\tau_b$, $\tau_c$ when $\tau_p > (\tau_{a} + \tau_{b} + \tau_{c})$.
\vspace{2mm}
\subsubsection{\textbf{WCET of dependent tasks}}
When a task is dependent on one or multiple tasks and implemented in different partitions under the same or different processor(s), the scheduling of the partitions can influence the response time, e.g., when an application is driven by two tasks $a$ and $b$, the WCET time of the application can be affected by the mutual arrangement of the tasks in partitions. 
Two tasks can be segregated in three possible ways; \textit{Case 1:} the tasks share the same partition, \textit{Case 2:} the tasks are implemented in different partitions, but under the same processor,  and \textit{Case 3:} the tasks are implemented in different processors.

\begin{figure}[h]
\begin{center}
\includegraphics[scale= 0.55]{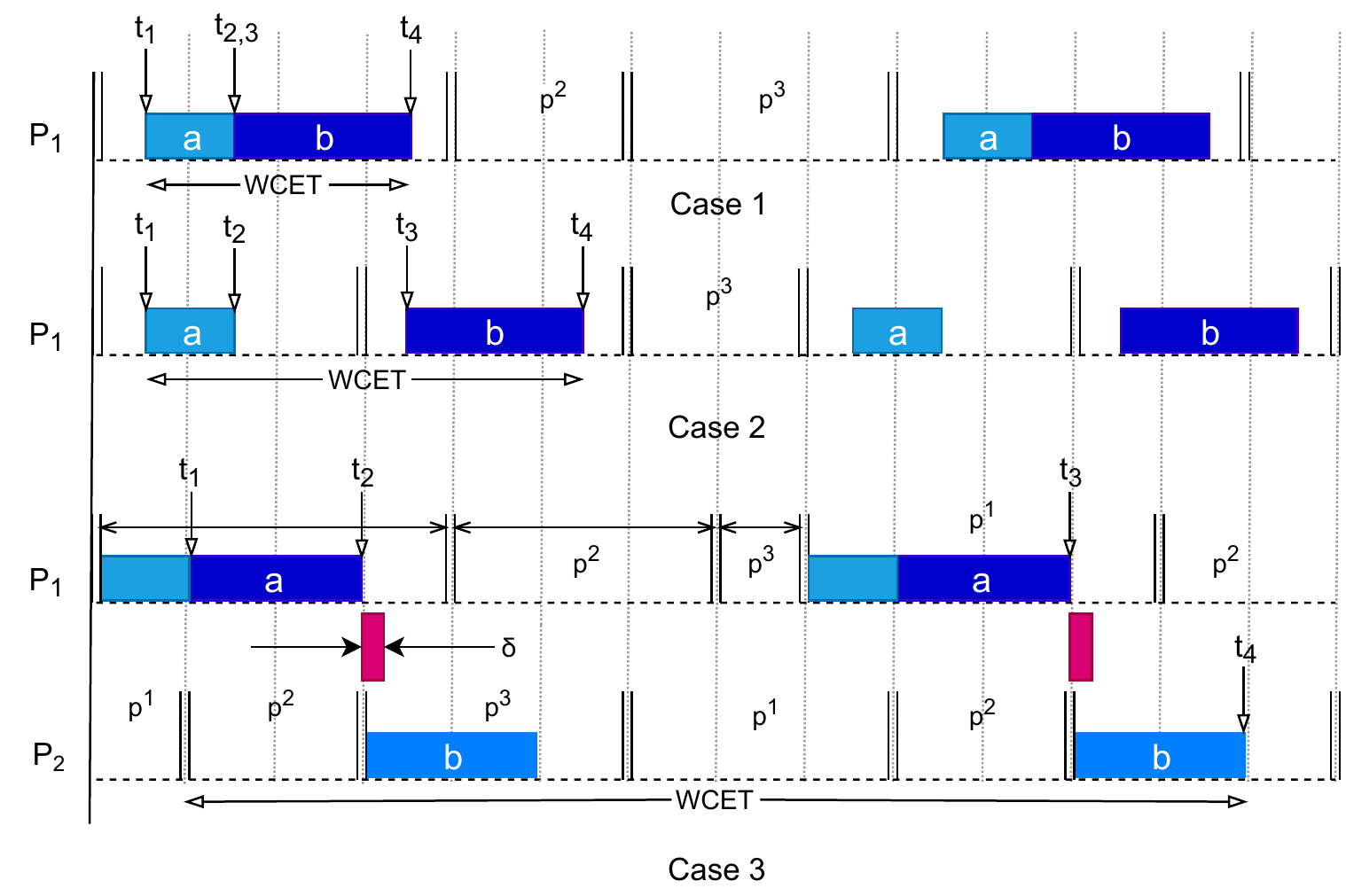}
\caption{A timing diagram showing WCET of a system with two sub-systems $a$ and $b$ under different implementations. Case 1: Both sub-systems are implemented in the same partition. Case 2: The sub-systems are implemented in different partitions on the same processor. Case 3: The sub-systems are implemented in separate processors, and $\delta$ is the worst-case end-to-end communication delay.}
\label{fig:wcet}
\end{center}
\end{figure}

\noindent \textit{Case 1:} When the tasks are scheduled sequentially within the same partition, there is no effect on WCET from the partition schedule, considering that the partition execution budget is higher than the collective WCET times of the tasks. The WCET of the application solely depends on the scheduling of the tasks inside the partition, and in this case $\tau_a + \tau_b$, as shown in Figure \ref{fig:wcet}.

\noindent \textit{Case 2:} When the tasks $a, b$ are scheduled at $\gamma_a, \gamma_b$ time respectively in two different partitions under the same processor, the WCET of the application $\tau_{s}$ is determined as:
\begin{equation*}
    \tau_{s} = (\tau_{p^1} - \gamma_a) + (\gamma_b + \tau_{p^2_b}) + \delta_{SO}
\end{equation*}
where, $\tau_{p^1}, \tau_{p^2}$ are execution budget of the partitions and $\delta_{SO}$ is partition switching overhead.   

\noindent \textit{Case 3:}
When the tasks $a,b$ are scheduled on different processors, the WCET of the application greatly depends on mutual scheduling of the partitions in two different processors. When, the partitions are synchronized, the best WCET of the application is:
\begin{equation*}
    WCET_{s} = (WCET_{p^1_a} + \delta_{CD} + WCET_{p^2_b})
\end{equation*}
where, $\delta_{CD}$ is the worst-case end-to-end communication delay. 
When there is no synchronization between the partitions in the processors and assumed to be unknown, the WCET of the system increases due to unpredictable relative state of computation in different partitions, and WCET of the application can be determined as:
\begin{equation*}
    WCET_{s} = \big(WCET_{p^1_a} + \tau_{p^1} + \tau_{p^2} + \tau_{p^3} + WCET_{p^3_b} - 1\big)
\end{equation*}
Figure \ref{fig:wcet} shows an example of the three possible distributions of the tasks $a,b$ and the impact of partition scheduling on WCET of the resulting functionality. 

\subsection{Task Synchronization}
Dependency in tasks can be mutual, i.e., two tasks $\alpha$ and $\beta$ can be inter-dependent. In such cases, synchronization must be established so that tasks can cooperate. For such synchronization, the processor features processor flags that are exposed to all the partitions under all processors, as explained earlier. A task can set the respective processor flags to some pre-set values as the execution progresses, and the other task can read the flag values and proceed with its execution path accordingly.

\begin{figure}[h!]
\begin{center}
\includegraphics[scale= 0.76]{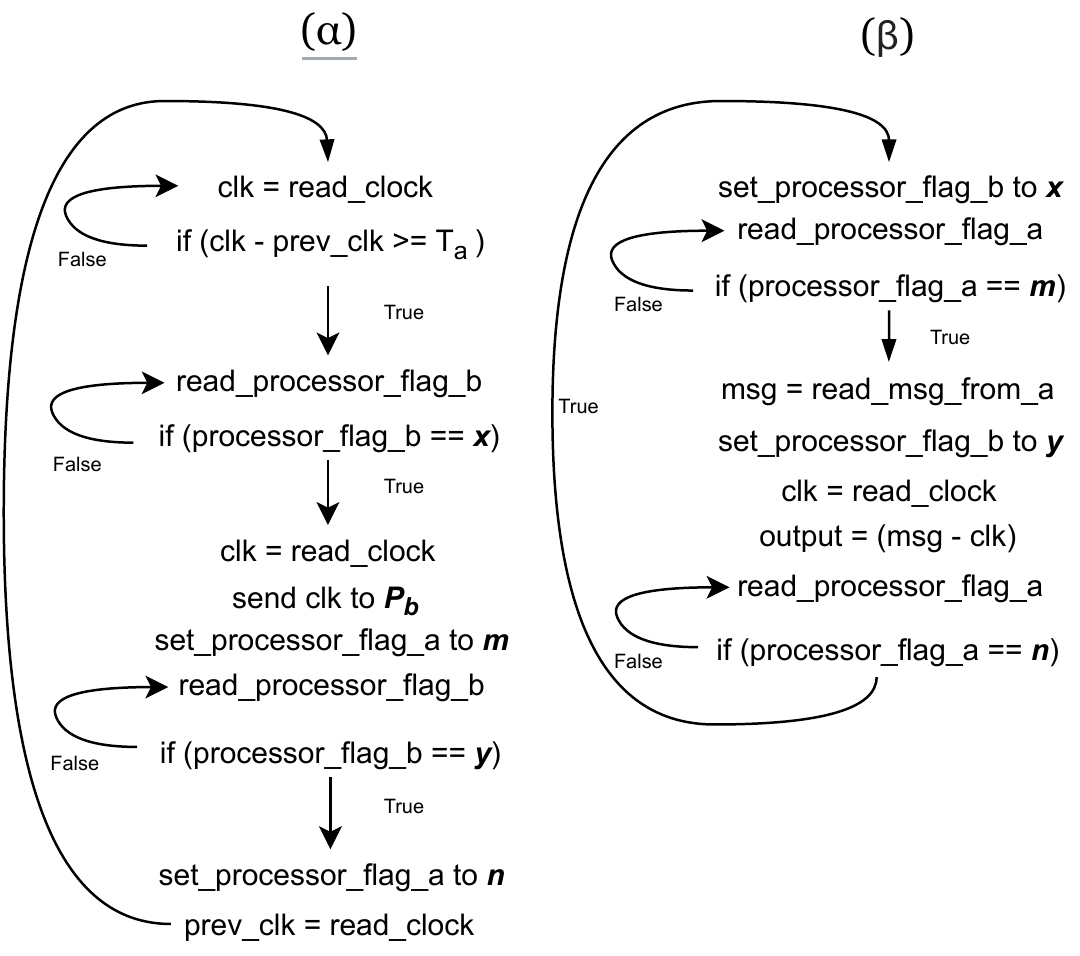}
\caption{Flow diagram showing synchronization process using processor flags between two tasks $\alpha$ and $\beta$ scheduled on different processors.}

\label{fig:exe}
\end{center}
\end{figure}

Figure \ref{fig:exe} demonstrates an example of tasks synchronization between the tasks $\alpha$ and $\beta$ implemented in different processors. 
The task $\beta$ sets associated flag to the value $x$ when it is ready. Task $\alpha$ reads the flag value and sends data to $\beta$. When the data is sent, $\alpha$ sets its flag to $m$. $\beta$ waits until the transmission is reflected in the flag controlled by $\alpha$. Task $\beta$ reads the data and sets its flag to $y$ to acknowledge the reception, and the process continues, as illustrated in Figure \ref{fig:exe}.

Note that the inter-system communication takes place over the NoC, where reception is acknowledged using the processor flags.  


\subsection{Reliability Analysis}
One of the motivations behind the proposed architecture is to eliminate any single point of failure and to enhance the reliability of the system. In the past, we have demonstrated a distributed implementation of a flight controller on an asymmetric multiprocessor to improve overall reliability \cite{majumderdalsgaard}.  

In \cite{majumderdalsgaard}, the reliability of the NoC is assumed to be extremely high (reliability 1.0), and also discussed that the NoC is a single point of failure in such architectures. However, a shared memory-based communication decreases the load on the NoC and can potentially prevent complete system failure while a failure in the NoC is encountered. Although, in this work, the shared memory does not feature any hardware-based protection, a software-based protection mechanism such as \textit{mutex} can be implemented.  

Single or multiple failures in the NI or in the router do not affect the rest of the network; however, the hub is a single point of failure in the proposed NoC architecture. The hub is not reachable by the executing software to modify the configuration and only susceptible to hardware failures. 

Each processor and even each partition can have separate memory devices (e.g., SRAM chips) for instruction and data memory, and thus erroneous operation of a memory device only affect the associated partition/ processor. For additional memory protection, redundant memory devices with a \textit{polling} mechanism can be implemented. 

Besides, PaRTAA features partition flags and processor flags that are visible to all other partitions in all processors. These flags can be strategically used to expose the execution state in each processor and the same can be observed by other systems in other partitions. This is a novel feature for health monitoring and safety-net mechanism, as well as task synchronization as it offers better insight over conventional \textit{time-out} implementations. The processor flags are memory-mapped and read-only, writing on these addresses has no effect, preventing any partition from altering the flags of other partitions. Lastly, such flags can potentially be used for communication as well (e.g., condition checks), thus, adding more to the communication redundancy. 

\subsection{System Configuration}
Each processor has isolated instruction memory, and the executable binary shall be downloaded to each instruction memory separately. Similarly, data needed for the execution shall be uploaded to each data memory of each partition under each processor individually. This could be a time-consuming process and may not be ideal for general purpose applications. However, once prepared, airborne systems are used for an extensive period of time without modification and that can justify the time-consuming configuration of the platform.

\section{Demonstration}
In our previous works, \cite{aero, majumder_nielsen}, we have demonstrated the timing and performance characteristics of the uniprocessor architecture and the NoC. Therefore, in this work, we have focused on demonstrating the feasibility and novelty of the proposed multiprocessor platform in the context of mixed-criticality airborne systems.

\subsection{Setup}
The proposed architecture is demonstrated on FPGA LEs. All the hardware is defined in \textit{Verilog} and synthesized with \textit{Intel Quartus Prime v18.1} tool. For experimentation and demonstration, we have used \textit{Intel Cyclone V SoC} board; however, the hard processor on the SoC is not used. The hardware platform features a 50 MHz oscillator, which is used to clock the processors, the NoC, and other co-processor IPs. No secondary clock or \textit{phase-locked loop} (PLL) is used. The FPGA resources utilized to synthesize the proposed architecture is presented in Table \ref{Tab:resource}.

\renewcommand{\arraystretch}{1.3}
\begin{table}[h!]
\begin{center}

\caption{Resource usage in FPGA synthesis.}
\label{Tab:resource}
\begin{tabular}{c|c|c|c}
\hline
\hline
 Hardware components & \begin{tabular}[c]{@{}c@{}}ALMs \\ \end{tabular}  & \begin{tabular}[c]{@{}c@{}} Combi. ALUTs  \\ \end{tabular}  & Registers           \\ \hline

\AE r\o \space Core        & 2653       & 2646    &  4493    \\ \hline
Co-processors       & 622            & 707     & 471                           \\ \hline
NoC       & 1862            & 2123    & 3647                           \\ \hline
\textbf{Total}       & \textbf{4137}            & \textbf{5476}    & \textbf{8611}                           \\ \hline
\AE r\o \space without partition       & 1607            & 1408    & 2277                           \\ \hline
Partition overhead        & 1046            & 1223    & 2216                           \\ \hline
\end{tabular}
\end{center}
\end{table}
\renewcommand{\arraystretch}{1}

We did not use any \textit{flash} memory to store the configuration, and the FPGA needs to be reconfigured after each power cycle. For direct volatile configuration, the synthesized configuration file (Intel's proprietary SOF format) is uploaded via a on-board \textit{JTAG} connection. The hardware platform is powered by an external power source, and the JTAG connection can be removed once the configuration file is downloaded on the FPGA. 
A specific GPIO pin is shorted that forces all four active-partition-flags from four processors to $00$, deactivating execution in all processors. 
Instructions for each processor and data for each partition is downloaded separately with two separate \textit{UART} connection (for data and instruction). This process is manual. Once all the data and instructions are downloaded, the GPIO pin is released, and a global reset button is pressed to start execution. The global reset clears all control and data registers and sets program counters to zero.

\renewcommand{\arraystretch}{1.2}
\begin{table*}[]
\begin{center}
\caption{Implemented systems with different criticality-levels and dependencies.}
\label{Tab:systems}
\begin{tabular}{cccccccc}
\hline \hline
\multirow{2}{*}{System}                                & \multirow{2}{*}{Criticality }                            & \multirow{2}{*}{Dependency}              & \multirow{2}{*}{Signal}               & \multirow{2}{*}{Processor}              & \multirow{2}{*}{Partition}              & \multirow{2}{*}{System WCET}                                   & \multirow{2}{*}{Partition WCET}      \\
                                                      &         &                                          &                                       &                                         &                                         & $\tau$ (ms)                                  & (ms)                \\ \hline
\multicolumn{1}{c|}{Flight Director}                   & \multicolumn{1}{c|}{1}                  & \multicolumn{1}{c|}{None}                & \multicolumn{1}{c|}{N/A}              & \multicolumn{1}{c|}{1}                  & \multicolumn{1}{c|}{1}                  & \multicolumn{1}{c|}{1.127}                  & 2                  \\ \hline
\multicolumn{1}{c|}{EIS}          & \multicolumn{1}{c|}{1}                  & \multicolumn{1}{c|}{Autopilot}           & \multicolumn{1}{c|}{Throttle command} & \multicolumn{1}{c|}{2}                  & \multicolumn{1}{c|}{1}                  & \multicolumn{1}{c|}{0.527}                  & 1                  \\ \hline
\multicolumn{1}{c|}{Moving Map}                        & \multicolumn{1}{c|}{3}                  & \multicolumn{1}{c|}{None}                & \multicolumn{1}{c|}{N/A}          & \multicolumn{1}{c|}{2}                  & \multicolumn{1}{c|}{3}                  & \multicolumn{1}{c|}{0.319}                  & 1                  \\ \hline
\multicolumn{1}{c|}{\multirow{2}{*}{Autopilot System}} & \multicolumn{1}{c|}{\multirow{2}{*}{2}} & \multicolumn{1}{c|}{\multirow{2}{*}{FD}} & \multicolumn{1}{c|}{Pitch Cue, Roll Cue}        & \multicolumn{1}{c|}{\multirow{2}{*}{1}} & \multicolumn{1}{c|}{\multirow{2}{*}{2}} & \multicolumn{1}{c|}{\multirow{2}{*}{1.761}} & \multirow{2}{*}{2} \\
\multicolumn{1}{c|}{}                                  & \multicolumn{1}{c|}{}                   & \multicolumn{1}{c|}{}                    & \multicolumn{1}{c|}{Speed (IAS), Ref Speed}         & \multicolumn{1}{c|}{}                   & \multicolumn{1}{c|}{}                   & \multicolumn{1}{c|}{}                   &                     \\ \hline
\end{tabular}%
\end{center}
\end{table*}
\renewcommand{\arraystretch}{1.0}

\subsection{Software Development}
The custom ISA used in this work also requires a custom compiler, which is an extensive work. On the other hand, writing assembly code is difficult and normally quite time consuming. Instead of developing an ISA specific compiler, we have used the standard \textit{x86 LLVM clang} compiler to generate assembly code, and we have developed an assembler to modify the x86 assembly to fit our ISA, before generating binary executables. We do not claim any novelty of this method in software development for the airborne system, but instead we demonstrate the competence of the proposed architecture with standard language like \textit{1999 ISO C} when software development tools are made available.

The register-register ISA prevents the use of \textit{immediate} and the software development must comply with it. All the variables and constants must be defined and initialized before use. This is not a challenge in the intended application as the software development guidelines for airborne systems require definition and initialization of all variables and constants.

\subsection{Avionics Use Case}
The avionics use case demonstrates the feasibility of porting a conventional avionics system onto the proposed architecture as well as its competence towards the requirements for a mixed-criticality system for airborne platforms. 
We have selected four generic avionics systems with different criticality-levels. The systems, dependencies, and communication requirements are presented in Table \ref{Tab:systems}. 

A \textit{flight-director} (FD) system, a DAL level A system (criticality-level 1 in our system), takes input data related to flight-path and flight data (heading, altitude, attitude, air-speed etc.) and computes the required manoeuvres to achieve a pre-set flight-path. The output of the FD is displayed as \textit{cues} on \textit{horizontal-situation-indicator} (HSI) on a \textit{primary-flight-display} (PFD) to guide the pilot or the autopilot system. 
An autopilot system can generate control commands to achieve reference altitude, heading, and speed based on inputs from an FD or a pilot. 
An \textit{engine-indicating-system} (EIS) displays the performance and health characteristics of the propulsion system on the \textit{electronic-flight-instrument-system} (EFIS) for crew awareness.
Lastly, a \textit{moving-map} system displays the aircraft's position on a variety of stored maps for crew awareness.

\begin{figure}[h]
\begin{center}
\includegraphics[scale= 0.5]{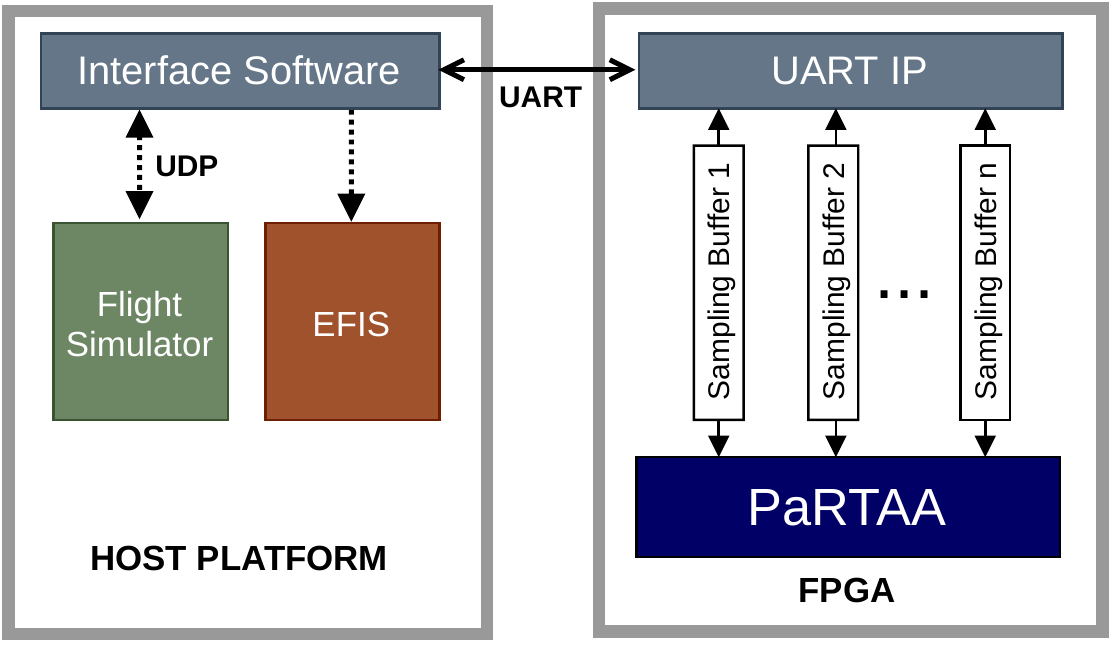} 
\caption{The avionics use case demonstration setup. The host system is a Linux platform that is connected to the FPGA experimental platform with a UART communication port.}
\label{fig:setup}
\end{center}
\end{figure}

For demonstration, we have developed each system using \textit{model-based-development} (MBD) technique with \textit{MATHWORKS Simulink}, where a mathematical/ logical model of the system is defined by the designer, and later C code is automatically generated from the model by the embedded code generator featured on Simulink tool. MBD is a well-adopted development technique in the aerospace industry, and the development process can potentially qualify to DO-178B/C \cite{DO178B, DO178C} requirements when the guidelines are followed. 

\begin{figure}[h]
\begin{center}
\includegraphics[scale= 0.45]{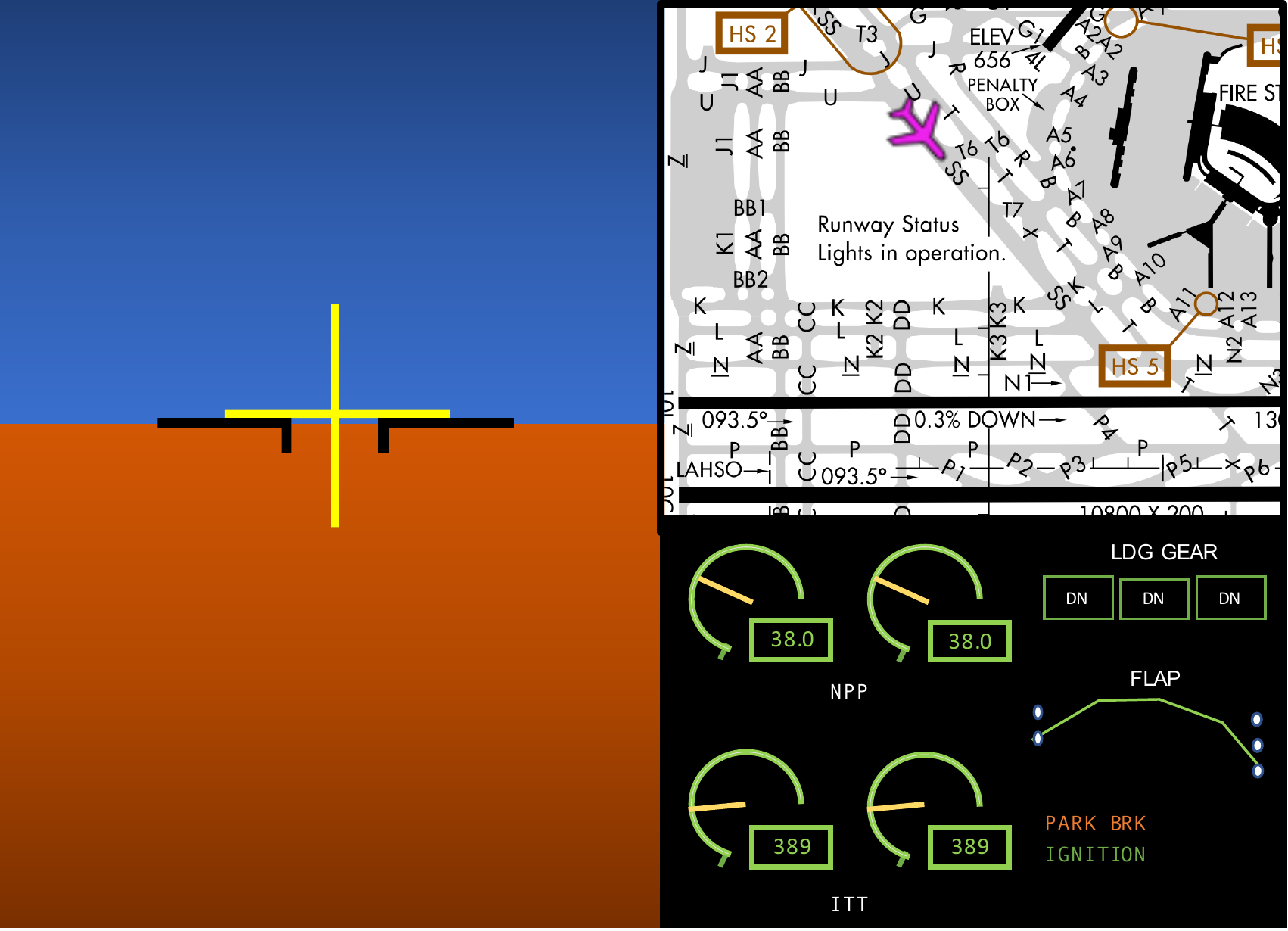}
\caption{A screenshot of an Electronic Flight Instrument System simulation with an \textit{Horizental Situation Indicator} application on the left, moving-map application on top-right and EIS application on bottom-right. }
\label{fig:mfd}
\end{center}
\end{figure}

To simulate the dynamics of the airborne platform and generate required data sources to evaluate the execution of the systems, we have used \textit{X-Plane 11 Flight Simulator}, a matured and well-adopted flight simulator used in several pilot training programs. 
The X-Plane 11 simulator executes on a host system (generic Linux platform) and features I/O interfacing (simulated sensor readings and control commands) over UDP. To establish communication between the simulator and the processor platform synthesized on the FPGA, we have used an \textit{interface software} that converts UDP packets from the simulator to UART and vice versa and features a UART communication port to interface with the FPGA platform. The demonstration setup is presented in Figure \ref{fig:setup}. 
Further, to visualize the computational outputs from the executing systems, we used an application to simulate an \textit{electronic flight instrument system} (EFIS), as shown in Figure \ref{fig:mfd}. Note that the multi-functional display (MFD) application works as a widget-based cockpit display system as defined in ARINC-661 \cite{ARINC661}, where only the widgets are driven by the input signals from the FPGA (via interface software), and the host system drives graphics. 

On the FPGA platform, a custom UART IP core receives the UART packets from the interfacing application and stores different signal values in separate sampling buffers, thus, preventing the received packets from being overwritten before consumption. This feature is essential to guarantee the availability of the received data packet for the correct consumer that may not have execution access at the time of the reception of the packet, which could be otherwise overwritten by another data-packet from a different signal or consumed by another system under partition with execution access. Such a sampling port-based UART is justifiable in this system as it mimics the behavior of an \textit{avionics-data-concentrator} unit used in an avionics system that provides isolation between data samples of different signals.

For demonstration, we have mapped the applications in different partitions and processors based on its criticality-levels, as presented in Table \ref{Tab:systems}. 
The FD system receives input signals from the simulator via UART sampling ports. 
For computational purpose, the FD uses only the protected data memory allocated to the partition. 
The output signals from the FD is sent to the MFD application via UART, and the output data is written to the shared memory associated with the processor. 
The autopilot system and the FD shares the same processor but executes in different partitions. The autopilot system reads the FD cues from the shared memory and uses protected memory for computation. The output is sent to the X-Plane simulator via UART. 
The EIS system is placed on a different processor and receives inputs from the simulator via UART, and from the autopilot system via the NoC. When the autopilot is engaged, the EIS reads the throttle\_command signal from NoC. The engagement/ disengagement of the autopilot is controlled by a processor flag. The computed output signals are sent to the MFD application using the UART port. 
The moving map application is independent.
The WCET of each system and partition is presented in Table \ref{Tab:systems}.

A dysfunction/ failure in the FD system directly affects the autopilot system due to its dependency on the FD output signals, irrespective of the communication medium. However, this has no effect on the systems independent of the FD system e.g., EIS and Moving map. 
A dysfunction in the autopilot system (DAL B) may corrupt the shared memory, which has no effect on the FD system (DAL A), as the FD system only writes data to the shared memory. 
The NoC is not a single point of failure in this architecture as a failure in the NoC does not affect the functionality of the FD and the autopilot system in spite of having inter-system dependencies.

\subsection{Comparison and Limitations}
Conventionally, the applications can be implemented in a mixed-criticality system with support from a hypervisor for establishing isolation between the applications with different criticality-levels. However, mapping the applications on PaRTAA does not require a hypervisor or host software support for establishing isolation, essential for mixed-criticality implementation.  
\begin{figure}[h]
\begin{center}
\includegraphics[scale= 0.3]{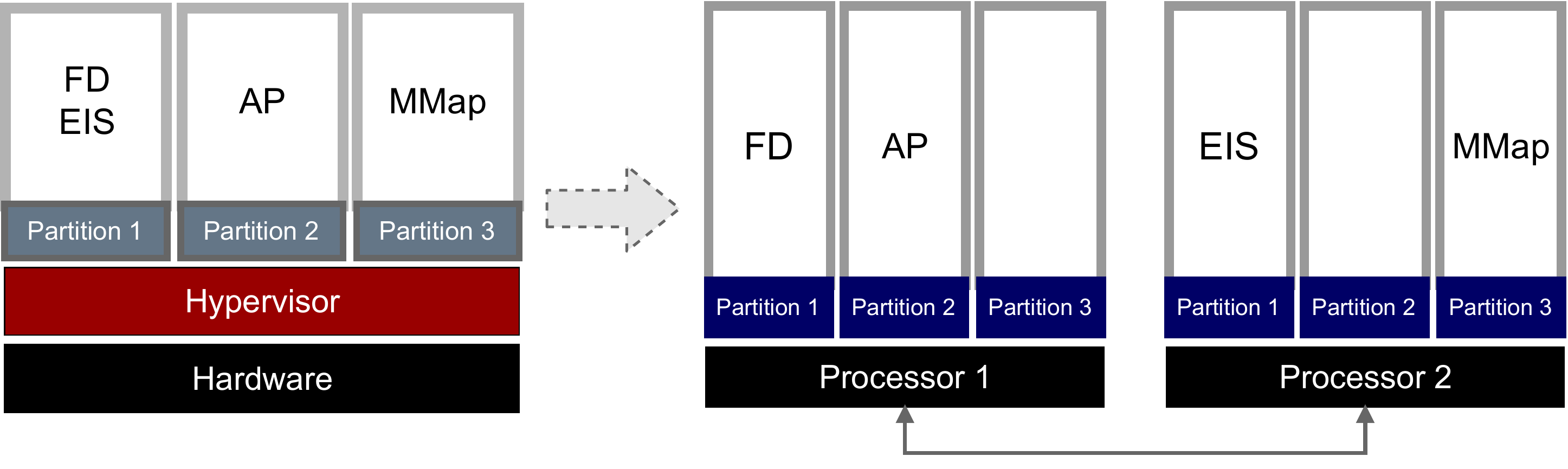}
\caption{Mapping of ARINC 653 partitions on the proposed architecture without any hypervisor.}
\label{fig:arinc}
\end{center}
\end{figure}
The ARINC 653 partition environment on a uniprocessor can accommodate multiple applications of the same criticality-level in the same partition; however only one application can execute at a time. Similarly, when applications are segregated on different processors on a single core equivalent multicore system, only a single application of a criticality-level gets execution access. In contrast, the proposed architecture enables the simultaneous execution of multiple applications of the same criticality-level. An example of potential mapping of ARINC 653 partitions on the proposed architecture is illustrated in Figure \ref{fig:arinc}.
Applications with high dependency can be implemented on the same processor 
(either on the same partition or different partition depending on the criticality-level), and data can be shared using either shared memory or using the NoC or both (for redundancy), e.g., the flight director and autopilot application share the same processor in the demonstration setup, which enhances the system reliability by removing a single point of failure.

Unlike asymmetric multicore or many-core architectures, where each system is mapped to a separate processor, this architecture features a more resource-efficient approach by allowing systems with different criticality-levels to share a processor while preserving isolation. Such implementation avails shared memory for data sharing and reduce the bandwidth requirement of inter-processor communication mechanism, such as a NoC.
On the other hand, the number of partitions in each processor is not largely scalable as only one partition gets execution access at any point of time, and increment in the number of partitions linearly decreases the periodicity of other partitions. However, there are only five levels of criticality defined for airborne systems, and level E systems are not implemented with other critical systems, which limits the requirement of partition to 4. However, the number of processor in the proposed architecture is scalable, and multiprocessor platform with more processors can be fabricated.

The second limitation comes from the IO interfacing as IOs are not synchronized with software execution. A data-packet can arrive at the destination when the consumer system may not have execution access. The received data-packet may get overwritten by another packet before the intended system gets access to execute, which limits the use of FIFO based and shared buffer system for IO interfacing, and all individual IOs shall have a dedicated sampling buffer. However, data-concentrator based avionics communication features sampling ports for data-protection and isolation. 
There is no specific requirement or constraint on the data or instruction memory interfaced with PaRTAA. However, the limited number of available GPIOs on the physical chip can be the bottleneck for interfacing large memory devices.

Lastly, the system scheduling on this architecture is complex as compared to single-core or symmetric-multicore scheduling due to the additional complexity of partition scheduling. Each system has a WCET and executes under a partition that has a direct influence on the WCET. 
For optimal scheduling problem, WCET of each system, communication delay between systems as well as partition execution times shall be taken into consideration. 

\section{Conclusion and Future Work}
The hardware-based isolation for mixed-criticality systems must enforce isolation in the temporal or the spatial domain to prevent interference between systems and often achieved with a large number of isolated processors that typically results in under-utilized hardware resources. The proposed multiprocessor architecture features coarse granularity within each processor for better utilization of resources while preserving isolation between partitions. Systems with intense inter-system communication requirements can share the same processor where shared memory can be used for data sharing, while systems with lower inter-system communication needs can be implemented on separate processors and communicate over the NoC. 

The proposed architecture unveils interesting and new scheduling challenges. Although we have successfully scheduled multiple avionics applications for experimentation and evaluation purposes, the scope of optimal partition scheduling, along with task scheduling within each partition remains unaddressed. Furthermore, the work can be investigated for reliability analysis, where the system can preserve critical functionality in the event of one or more subsystem failure. 

\section{Acknowledgement}
We would like to thank Prof. Edward Lee from UC, Berkeley and Prof. Peter Koch from Aalborg University, Denmark for their insightful comments and feedback.

This research is funded by the Danish Independent Research Foundation under the Ministry of Higher Education and Science, Denmark under grant number 
6111-00363B.
\ifCLASSOPTIONcaptionsoff
  \newpage
\fi

\bibliography{bibfile}

\begin{thebibliography}{10}
\providecommand{\url}[1]{#1}
\csname url@samestyle\endcsname
\providecommand{\newblock}{\relax}
\providecommand{\bibinfo}[2]{#2}
\providecommand{\BIBentrySTDinterwordspacing}{\spaceskip=0pt\relax}
\providecommand{\BIBentryALTinterwordstretchfactor}{4}
\providecommand{\BIBentryALTinterwordspacing}{\spaceskip=\fontdimen2\font plus
\BIBentryALTinterwordstretchfactor\fontdimen3\font minus
  \fontdimen4\font\relax}
\providecommand{\BIBforeignlanguage}[2]{{%
\expandafter\ifx\csname l@#1\endcsname\relax
\typeout{** WARNING: IEEEtran.bst: No hyphenation pattern has been}%
\typeout{** loaded for the language `#1'. Using the pattern for}%
\typeout{** the default language instead.}%
\else
\language=\csname l@#1\endcsname
\fi
#2}}
\providecommand{\BIBdecl}{\relax}
\BIBdecl

\bibitem{DO297}
{Radio Technical Commission for Aeronautics}, \emph{RTCA/DO-297: Integrated
  Modular Avionics (IMA) Development Guidance and Certification
  Considerations}, 2005.

\bibitem{LuiSha}
L.~Sha, M.~Caccamo, R.~Mancuso, J.-E. Kim, M.-K. Yoon, R.~Pellizzoni, H.~Yun,
  R.~Kegley, D.~Perlman, G.~Arundale, and R.~Bradford, ``Single core equivalent
  virtual machines for hard real—time computing on multicore processors,''
  2014.

\bibitem{aero}
S.~{Majumder}, J.~F.~D. {Nielsen}, and T.~{Bak}, ``{\AE}r\o: A platform
  architecture for mixed-criticality airborne systems,'' \emph{IEEE
  Transactions on Computer-Aided Design of Integrated Circuits and Systems},
  pp. 1--1, 2019.

\bibitem{majumder_nielsen}
S.~Majumder, J.~Nielsen, A.~La~Cour-Harbo, H.~Schi{\o}ler, and T.~Bak, ``A
  real-time on-chip network architecture for mixed criticality aerospace
  systems,'' \emph{The Aeronautical Journal}, vol. 123, no. 1269, p.
  1788–1806, 2019.

\bibitem{DO178B}
{Radio Technical Commission for Aeronautics}, \emph{RTCA/DO-178B: Software
  Considerations in Airborne Systems and Equipment Certification}, 1992.

\bibitem{DO254}
{Radio Technical Commission for Aeronautics}, \emph{RTCA/DO-254: Design
  Assurance Guidance for Airborne Electronic Hardware}, 2000.

\bibitem{DO178C}
{Radio Technical Commission for Aeronautics}, \emph{RTCA/DO-178C: Software
  Considerations in Airborne Systems and Equipment Certification}, December
  2011.

\bibitem{CAST32A}
``Certification authorities software team: Position paper cast-32a, multi-core
  processors,'' 2016.

\bibitem{DO248C}
{Radio Technical Commission for Aeronautics}, \emph{RTCA/DO-248C: “Supporting
  Information for DO-178C and DO-278A.}, 2011.

\bibitem{ARINC653}
{Airlines electronic engineering committee (AEEC), ARINC, Inc.}, \emph{avionics
  application software standard interface (ARINC specification 653-1).}, 2003.

\bibitem{Mahapatra2011MicroprocessorEF}
R.~N. Mahapatra, J.~Lee, N.~Gupta, and B.~Manners, ``Federal aviation
  administration: Microprocessor evaluations for safety-critical, real-time
  applications: Authority for expenditure no. 43 phase 5 report,'' 2011.

\bibitem{Mahapatra2009MicroprocessorEF}
R.~N. Mahapatra, P.~Bhojwani, J.~H. Lee, and Y.~Kim, ``Federal aviation
  administration: Microprocessor evaluations for safety-critical, real-time
  applications: Authority for expenditure no. 43 phase 3 report,'' 2009.

\bibitem{rushby2000partitioning}
J.~Rushby, ``Partitioning in avionics architectures: Requirements, mechanisms,
  and assurance,'' National Aeronautics and Space Administration (NASA), Tech.
  Rep., 2000.

\bibitem{144077}
L.~{Lindh}, ``Fastchart-a fast time deterministic cpu and hardware based
  real-time-kernel,'' in \emph{Proceedings. EUROMICRO `91 Workshop on Real-Time
  Systems}, June 1991, pp. 36--40.

\bibitem{557849}
J.~{Adomat}, J.~{Furunas}, L.~{Lindh}, and J.~{Starner}, ``Real-time kernel in
  hardware rtu: a step towards deterministic and high-performance real-time
  systems,'' in \emph{Proceedings of the Eighth Euromicro Workshop on Real-Time
  Systems}, June 1996, pp. 164--168.

\bibitem{5567091}
T.~{Ungerer}, F.~{Cazorla}, P.~{Sainrat}, G.~{Bernat}, Z.~{Petrov},
  C.~{Rochange}, E.~{Quinones}, M.~{Gerdes}, M.~{Paolieri}, J.~{Wolf},
  H.~{Casse}, S.~{Uhrig}, I.~{Guliashvili}, M.~{Houston}, F.~{Kluge},
  S.~{Metzlaff}, and J.~{Mische}, ``Merasa: Multicore execution of hard
  real-time applications supporting analyzability,'' \emph{IEEE Micro},
  vol.~30, Sep. 2010.

\bibitem{6925994}
M.~{Zimmer}, D.~{Broman}, C.~{Shaver}, and E.~A. {Lee}, ``Flexpret: A processor
  platform for mixed-criticality systems,'' in \emph{2014 IEEE 19th Real-Time
  and Embedded Technology and Applications Symposium (RTAS)}, April 2014, pp.
  101--110.

\bibitem{6378622}
I.~{Liu}, J.~{Reineke}, D.~{Broman}, M.~{Zimmer}, and E.~A. {Lee}, ``A pret
  microarchitecture implementation with repeatable timing and competitive
  performance,'' in \emph{2012 IEEE 30th International Conference on Computer
  Design (ICCD)}, Sep. 2012, pp. 87--93.

\bibitem{6341002}
D.~{May}, ``The xmos architecture and xs1 chips,'' \emph{IEEE Micro}, vol.~32,
  no.~6, pp. 28--37, Nov 2012.

\bibitem{1212740}
M.~{Delvai}, W.~{Huber}, P.~{Puschner}, and A.~{Steininger}, ``Processor
  support for temporal predictability - the spear design example,'' in
  \emph{15th Euromicro Conference on Real-Time Systems, 2003. Proceedings.},
  July 2003, pp. 169--176.

\bibitem{ELSALLOUM20131020}
C.~E. Salloum], M.~Elshuber, O.~Höftberger, H.~Isakovic, and A.~Wasicek, ``The
  across mpsoc – a new generation of multi-core processors designed for
  safety–critical embedded systems,'' \emph{Microprocessors and
  Microsystems}, vol.~37, no. 8, Part C, pp. 1020 -- 1032, 2013.

\bibitem{TRUJILLO2014921}
S.~Trujillo, A.~Crespo, A.~Alonso, and J.~Pérez, ``Multipartes: Multi-core
  partitioning and virtualization for easing the certification of
  mixed-criticality systems,'' \emph{Microprocessors and Microsystems},
  vol.~38, no. 8, Part B, pp. 921 -- 932, 2014.

\bibitem{PEREZ2017145}
H.~Pérez, J.~J. Gutiérrez, S.~Peiró, and A.~Crespo, ``Distributed
  architecture for developing mixed-criticality systems in multi-core
  platforms,'' \emph{Journal of Systems and Software}, vol. 123, 2017.

\bibitem{baruah2015mixed}
S.~K. Baruah, L.~Cucu-Grosjean, R.~I. Davis, and C.~Maiza, ``Mixed criticality
  on multicore/manycore platforms (dagstuhl seminar 15121),'' in \emph{Dagstuhl
  Reports}, vol.~5, no.~3.\hskip 1em plus 0.5em minus 0.4em\relax Schloss
  Dagstuhl-Leibniz-Zentrum fuer Informatik, 2015.

\bibitem{Sano2015}
K.~Sano, D.~Soudris, M.~H{\"{u}}bner, and P.~C. Diniz, ``{Applied
  reconfigurable computing 11th International symposium, ARC 2015 Bochum,
  Germany, april 13-17, 2015 proceedings},'' \emph{Lecture Notes in Computer
  Science (including subseries Lecture Notes in Artificial Intelligence and
  Lecture Notes in Bioinformatics)}, vol. 9040, pp. 191--201, 2015.

\bibitem{16}
D.~Wiklund and D.~Liu, ``Socbus: switched network on chip for hard real time
  embedded systems,'' in \emph{Proceedings International Parallel and
  Distributed Processing Symposium}, April 2003.

\bibitem{15}
D.~Bertozzi and L.~Benini, ``Xpipes: a network-on-chip architecture for
  gigascale systems-on-chip,'' \emph{IEEE Circuits and Systems Magazine},
  vol.~4, pp. 18--31, 2004.

\bibitem{18}
P.~H. Pham, J.~Park, P.~Mau, and C.~Kim, ``Design and implementation of
  backtracking wave-pipeline switch to support guaranteed throughput in
  network-on-chip,'' \emph{IEEE Transactions on Very Large Scale Integration
  (VLSI) Systems}, vol.~20, no.~2, pp. 270--283, Feb 2012.

\bibitem{17}
P.~T. Wolkotte, G.~J.~M. Smit, G.~K. Rauwerda, and L.~T. Smit, ``An
  energy-efficient reconfigurable circuit-switched network-on-chip,'' in
  \emph{19th IEEE International Parallel and Distributed Processing Symposium},
  April 2005, pp. 155a--155a.

\bibitem{26}
E.~Bolotin, I.~Cidon, R.~Ginosar, and A.~Kolodny, ``Qnoc: Qos architecture and
  design process for network on chip,'' \emph{Journal of Systems Architecture},
  vol.~50, no.~2, pp. 105 -- 128, 2004, special issue on networks on chip.

\bibitem{27}
S.~H. Lo, Y.~C. Lan, H.~H. Yeh, W.~C. Tsai, Y.~H. Hu, and S.~J. Chen, ``Qos
  aware binoc architecture,'' in \emph{2010 IEEE International Symposium on
  Parallel Distributed Processing (IPDPS)}, April 2010, pp. 1--10.

\bibitem{28}
E.~d.~F. Corr\^{e}a, L.~A. d. P.~e. Silva, F.~R. Wagner, and L.~Carro,
  ``Fitting the router characteristics in nocs to meet qos requirements,'' in
  \emph{Proceedings of the 20th Annual Conference on Integrated Circuits and
  Systems Design}, ser. SBCCI '07.\hskip 1em plus 0.5em minus 0.4em\relax New
  York, NY, USA: ACM, 2007, pp. 105--110.

\bibitem{29}
C.~H. Lu, K.~C. Chiang, and P.~A. Hsiung, ``Round-based priority arbitration
  for predictable and reconfigurable network-on-chip,'' in \emph{2009
  International Conference on Field-Programmable Technology}, Dec 2009, pp.
  403--406.

\bibitem{MULCORS}
X.~JEAN, M.~G.~G. BERTHON, and M.~FUMEY, ``Mulcors: Use of multicore procesors
  in airborne systems, easa,'' 2011.

\bibitem{Lo}
M.~Lo, N.~Valot, F.~Maraninchi, and P.~Raymond, ``{IMPLEMENTING A REAL-TIME
  AVIONIC APPLICATION ON A MANY-CORE PROCESSOR},'' in \emph{{42nd European
  Rotorcraft Forum (ERF)}}, Lille, France, Sep. 2016.

\bibitem{SCHOEBERL2015449}
M.~Schoeberl, S.~Abbaspour, B.~Akesson, N.~Audsley, R.~Capasso, J.~Garside,
  K.~Goossens, S.~Goossens, S.~Hansen, R.~Heckmann, S.~Hepp, B.~Huber,
  A.~Jordan, E.~Kasapaki, J.~Knoop, Y.~Li, D.~Prokesch, W.~Puffitsch,
  P.~Puschner, A.~Rocha, C.~Silva, J.~Sparsø, and A.~Tocchi, ``T-crest:
  Time-predictable multi-core architecture for embedded systems,''
  \emph{Journal of Systems Architecture}, vol.~61, no.~9, pp. 449 -- 471, 2015.

\bibitem{7445422}
A.~{Rocha}, C.~{Silva}, R.~B. {Sørensen}, J.~{Sparsø}, and M.~{Schoeberl},
  ``Avionics applications on a time-predictable chip-multiprocessor,'' in
  \emph{2016 24th Euromicro International Conference on Parallel, Distributed,
  and Network-Based Processing (PDP)}, Feb 2016, pp. 777--785.

\bibitem{majumderdalsgaard}
S.~Majumder, J.~F. Dalsgaard~Nielsen, T.~Bak, and A.~la~Cour-Harbo, ``Reliable
  flight control system architecture for agile airborne platforms: an
  asymmetric multiprocessing approach,'' \emph{The Aeronautical Journal}, vol.
  123, no. 1264, p. 840–862, 2019.

\bibitem{ARINC661}
{Airlines electronic engineering committee (AEEC), ARINC, Inc.}, \emph{ARINC
  661 COCKPIT DISPLAY SYSTEM INTERFACES TO USER SYSTEMS ARINC SPECIFICATION},
  2001.

\end{thebibliography}


\vskip -2\baselineskip plus -1fil
\begin{IEEEbiography}[{\includegraphics[width=1in,height=1.25in,clip,keepaspectratio]{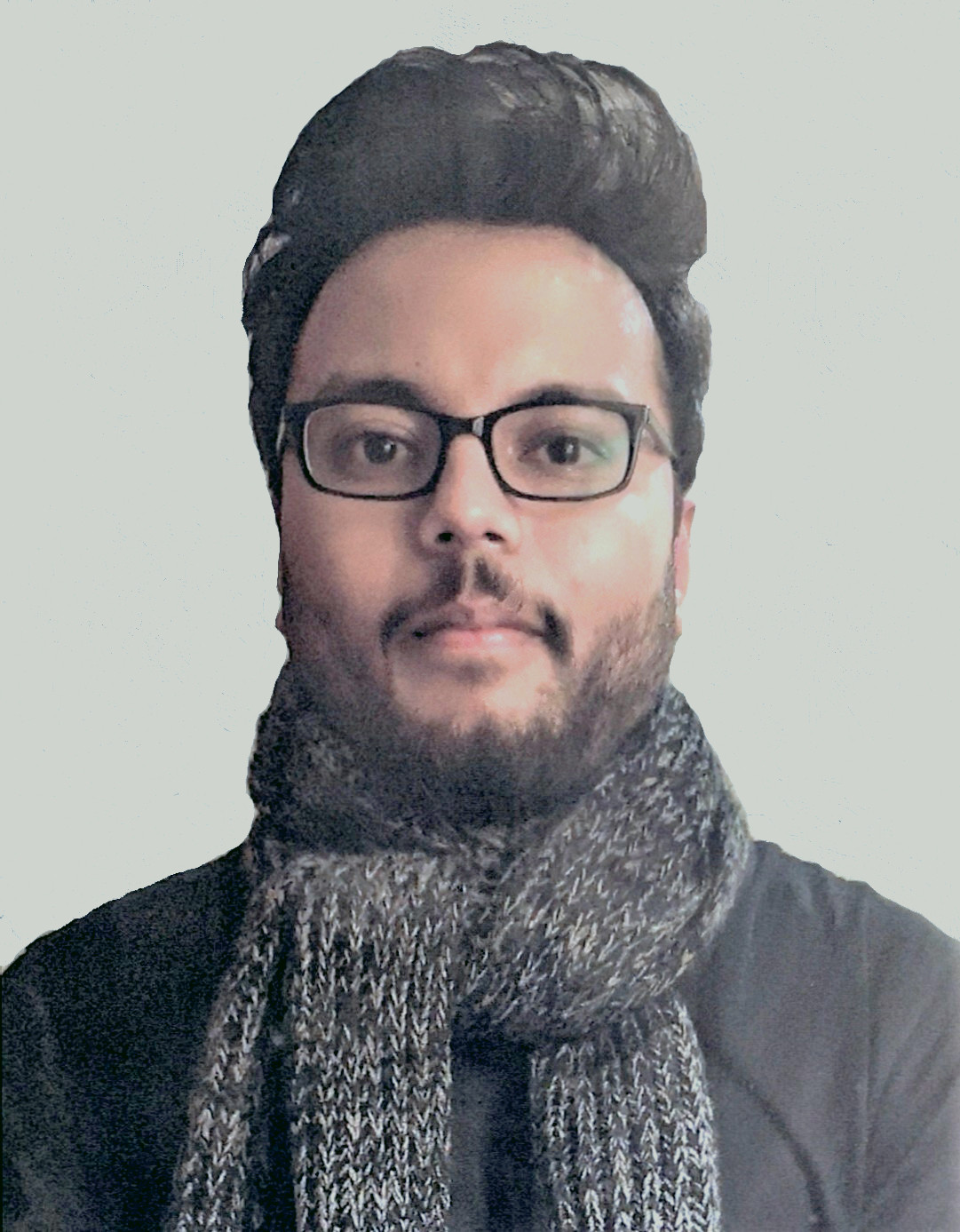}}]{Shibarchi Majumder} is a doctoral researcher at the Department of Electronic Systems, Aalborg University, Aalborg, Denmark. 
He has years of industrial and academic experience in avionics systems, embedded flight computing, hardware systems and has several publications in related domains. 
His research interests include real-time systems, mixed-criticality systems, hardware design, VLSI, embedded computation and unmanned aerial systems. Earlier, he
received a bachelor's degree in Aerospace Engineering and a master's degree in Avionics Engineering. 
\end{IEEEbiography}

\vskip -2\baselineskip plus -1fil 
\begin{IEEEbiography}[{\includegraphics[width=1in,height=1.25in,clip,keepaspectratio]{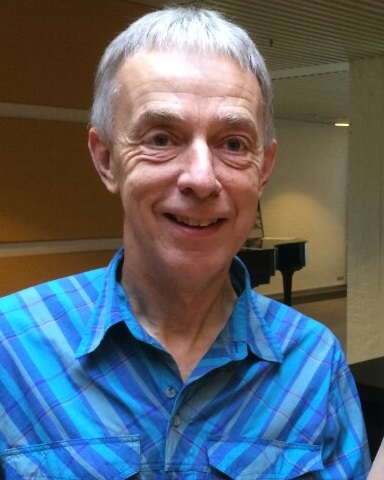}}]{Jens Frederik Dalsgaard Nielsen} is employed as Associate Professor at Aalborg University at the section Automation \& Control. He has a Master of Science in EE and a PhD within automation and control domain.
For more than 15 years he has been heading the student satellite activities at Aalborg University which has launched 5 cubesats 100\% developed at AAU and participated in three other launches. His primary domain is real-time systems ranging from hardware to real time operating systems, networking for safety critical systems and software development.
\end{IEEEbiography}

\vskip -2\baselineskip plus -1fil 

\begin{IEEEbiography}[{\includegraphics[width=1in,height=1.25in,clip,keepaspectratio]{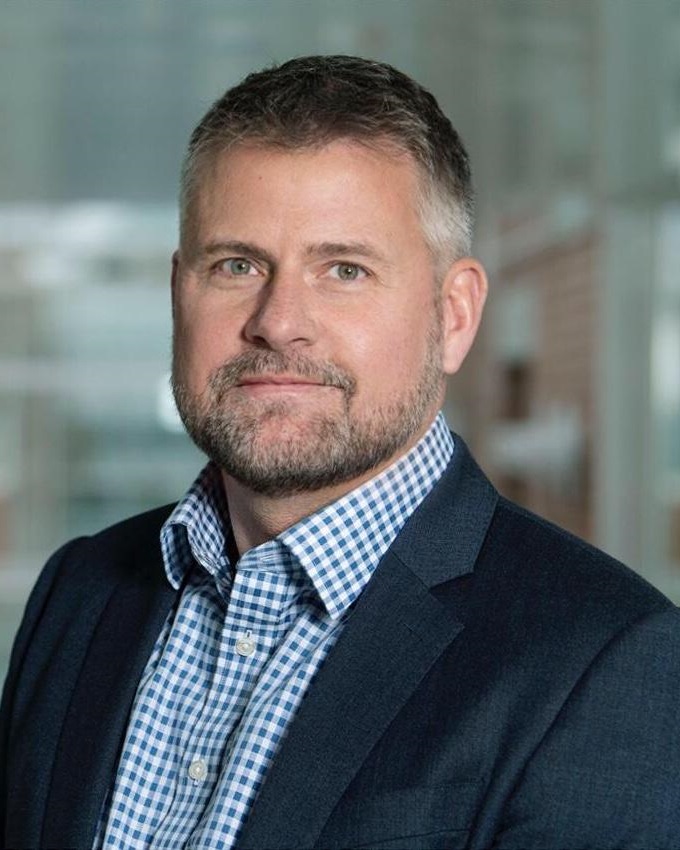}}]{Thomas Bak}received a PhD degree in control systems from Aalborg University in 1998. He became an Assistant Professor in 1998, an Associate Professor in 1999, and a Full Professor of autonomous systems in 2006. From 2003-2006 he was a senior researcher and head of research unit at Aarhus University. 
Since 2018 he is head of the Department of Electronic Systems. He has published more than 100 papers in the fields of control and its applications. His current research interest includes autonomous systems and robotics. He is a senior member of IEEE and chair of the IEEE Joint Chapter on Control System Society and Robotics and Automation Society. He is associated editor for the European Journal of Control.
\end{IEEEbiography}




\end{document}